# Tomography Scan of Charge Density Wave in NbSe$_2$

Jyun-Yu Wu[1], Yung-Ting Lee[2], Guan-Hao Chen[1,4], Zheng-Hong Li[1], Chang-Tsan Lee[1], Jie-Yu Hsu[1], Chia-Nung Kuo[3], Juhn-Jong Lin[1], Wen-Hao Chang[1,4], Chin-Shan Lue[3,5], Po-Tuan Cheng[2], Cheng-Tien Chiang[6,8], Chien-Cheng Kuo[7], Chien-Te Wu[1], Chi-Cheng Lee[8]*, Ming-Chiang Chung[9]*, Hung-Chung Hsueh[8]*, Chun-Liang Lin[1]*

[1]Department of Electrophysics, National Yang Ming Chiao Tung University, No. 1001 University Rd., Hsinchu 300, Taiwan

[2]Department of Vehicle Engineering, National Taipei University of Technology, No. 1, Sec. 3, Chung-Hsiao E. Rd, Taipei 106, Taiwan

[3]Department of Physics, National Cheng Kung University, No. 1 University Rd., Tainan 701, Taiwan

[4]Research Center for Applied Sciences, Academia Sinica, 128 Academia Road, Section 2, Nankang, Taipei 11529, Taiwan

[5]Consortium of Emergent Crystalline Materials, National Science and Technology Council, Taipei 106, Taiwan

[6]Institute of Atomic and Molecular Sciences, Academia Sinica, No. 1, Roosevelt Rd., Sec. 4, Taipei 106, Taiwan

[7]Department of Physics, National Sun Yet-Sen University, No.70 Lien-hai Rd., Kaohsiung 804, Taiwan, Taiwan

[8]Department of Physics, Tamkang University, No.151, Yingzhuan Rd., New Taipei City 251, Taiwan

[9]Department of Physics, National Chung Hsing University, No. 145 Xingda Rd., Taichung 402, Taiwan

## Abstract

Charge density wave (CDW) resulted from a small distortion in the lattice is able to create new orders beyond the original lattice. In *2H*-NbSe$_2$, one of the layered transition metal dichalcogenides (TMD), the 3×3 charge order appears in two-dimensional (2D) layers. Although CDW is usually described by a sine wave, the spatial distribution within a 2D layer has never been systematically visualized. Here by using scanning tunneling microscopy (STM) and density functional theory (DFT), we have monitored the evolution of 3×3 CDW along *c*-axis and realized a nearly tomography scan of CDW of the topmost layer. The results show that the strength of 3×3 charge order varies while increasing the tunneling current. The 3×3 charge order is relatively strong at the outermost Se level and decreases while probing in between Se and Nb levels. Interestingly, the 3×3 charge order gets strong again as reaching Nb level but along with a phase shift. We further calculated the orbital charge distributions and found that both CDW intensity modulation and phase shift are strongly correlated with the distribution of Se *p* orbitals and Nb *d* orbitals.

**Introduction**

Charge density waves (CDW) arising from Peierls transition can create a new order inside crystals [1]. Influenced by the new order, several macroscopic properties similar to superconductivities can be observed. For example, the nonlinear changes of resistivity in crystals are observed in temperature-dependent transport measurements [2, 3] and Fermi surface nesting is revealed by photoemission spectroscopy below transition temperature [4, 5]. In microscopic points of view, the new order of the charge density is strongly and spontaneously correlated with the periodic displacement in the lattice. Yet, the correlation between the charge density and lattice rearrangement still lacks a comprehensive experimental proof. To significantly clarify this correlation, a 2D lattice with CDW phase is preferred. Here, NbSe$_2$ is chosen because it is a layered TMDs and in each layer only contains two elements. For *2H*-NbSe$_2$, it shows 3×3 CDW below 35 K in bulk crystal [6] and $T_{CDW}$ can be further enhanced to 145 K for monolayer [7]. On the other hand, CDW and superconductivity can coexist in *2H*-NbSe$_2$ bulk up to 7.2 K [8]. A former study has visualized the layer-dependent superconducting properties of *2H*- NbSe$_2$ [9], but the distribution of CDW along c-axis has not yet been visualized.

Scanning tunneling microscopy (STM) is a suitable method to visualize CDW because of its sub-angstrom resolution. Not only for conventional CDW in layer materials [10-13], the relationship between unconventional charge orders or even pair density wave and lattice geometry in Kagome materials such as AV$_3$Sb$_5$ (A = K, Cs, Rb) can also be identified [14-16]. Many previous studies have shown the evolution of CDW by tuning the sample bias, i.e. the energy levels [17-20]. Here, by tuning the tunneling current, STM allows us to probe the distribution of charge density at different heights within the layer [21]. It is found that the intensity of the 3×3 charge order exhibits a strong correlation with the tunneling current, i.e. tip-sample distance, in both filled and hollow configurations. By comparing with the 1×1 atomic lattice, the 3×3 charge order gradually disappears with the tip approaching and appears again when the tip approaches more. It is an evidence to show that the distribution of the CDW is not homogeneous along *c*-axis as expected. A phase shift is found when probing the 3×3 charge order at Se and Nb levels. Similar results can be reproduced by the simulation based on density functional theory (DFT), further confirming that it can be regarded as a tomography scan of the 2D CDW.

**Method**

The single crystal *2H*-NbSe$_2$ was introduced into the UHV chamber at a base pressure of 2×10$^{-10}$ torr and *in situ* cleaved. The STM measurements were performed under liquid He conditions. STM tip was made from tungsten that was prepared by electrochemical etching in a 3M NaOH solution. Based on DFT, norm-conserving pseudopotentials, and optimized pseudoatomic basis functions, the first-principles calculations of the NbSe$_2$ materials were calculated by using the OpenMX (3.9.9) code [24-29]. The generalized gradient approximation (GGA) was used in the Perdew-Burke-Ernzerhof (PBE) scheme [30] to treat the exchange-correlation function. The optimized radial functions used were Nb-s3p2d2 and Se-s3p2d2 for niobium and selenium, where the abbreviations of basis functions represent (atomic symbol)-(number of radial functions for *s*, *p*, and *d* orbitals). In DFT calculations, the criterions of force convergence and electronic self-consistent field were $10^{-4}$ hartree/bohr and $10^{-8}$ hartree, respectively. The *k*-grid in the NbSe$_2$ 3×3×1 monolayers is 9×9×1. The energy cutoff was 600 (Ry). The Tersoff-Hamann scheme was employed to simulate STM images with partial charge density [31]. The NbSe$_2$ structures and charge density iso-surfaces are plotted by using VESTA [32] and OpenMX Viewer [33], respectively.

**Results and Discussion**

For the 3×3 CDW in *2H*-NbSe$_2$, the electron-phonon coupling can stabilize the atomic structure in several different configurations [6, 19, 22, 23]. The filled and hollow phases as described in **Figure 1** are two most-observed configurations in the experiments [6, 19]. The difference in these two configurations is mainly due to the various rearrangements of Nb atoms. **Figure 1 (a)** shows the STM image of an *in situ* cleaved *2H*-NbSe$_2$ surface and both the 1×1 atomic lattice and 3×3 charge order are observed. Similar to several previous researches [6, 19], two different phases can be found simultaneously. One is "hollow" phase as indicated by the green square and the other is "filled" as indicated by the blue square. In both phases, the distorted lattice caused by electron-phonon coupling results in the rearrangement of Nb atoms. As shown in **Figure 1 (b)** and **(c)**, triangular clustering of Nb atoms forms big (six Nb) and small (three Nb) triangles which are empty at the center for hollow phase but filled with a Se atom for filled phase. The difference can also be observed in STM images. At +10 mV sample bias, the hollow phase reveals bright three-pointed stars driven by triangular clustering of three Nb atoms as shown in **Figure 1(e)**. On the other hand, the filled phase reveals bright spots driven by triangular clustering of six Nb atoms as shown in **Figure 1(f)**. After the fast Fourier transform (FFT), both phases show primary 3×3 CDW wave-vectors presented together with the Bragg vectors of 1×1 atomic lattice as revealed in the insets of **Figure 1 (e)** and **(f)**. Interestingly, while tunneling current is increased from 1 nA to 100 nA, different apparent images can be obtained as shown by **Figure 1 (g)** and **(h)** for hollow and filled phases, respectively. It is known that the distance between the tip and the sample in the STM tunneling junction can be tuned by the set point of tunneling currents. Meanwhile it allows us to probe the corresponding real-space distribution of density of states (DOS) at different depth with respect to the outermost surface [21]. The higher value of the tunneling current, the deeper level is probed as illustrated in **Figure 1(d)**. Although the periodicity of both atomic lattices and charge orders remains the same as indicated by the insets in **Figure 1 (g)** and **(h)**, the intensity may vary as predicted in a previous research of a similar TMD material, NbS$_2$ [10]. Therefore, a systematic investigation is required to understand the evolution of 3×3 charge order along *c*-axis from Se level to Nb level by tuning the set points of tunneling current.

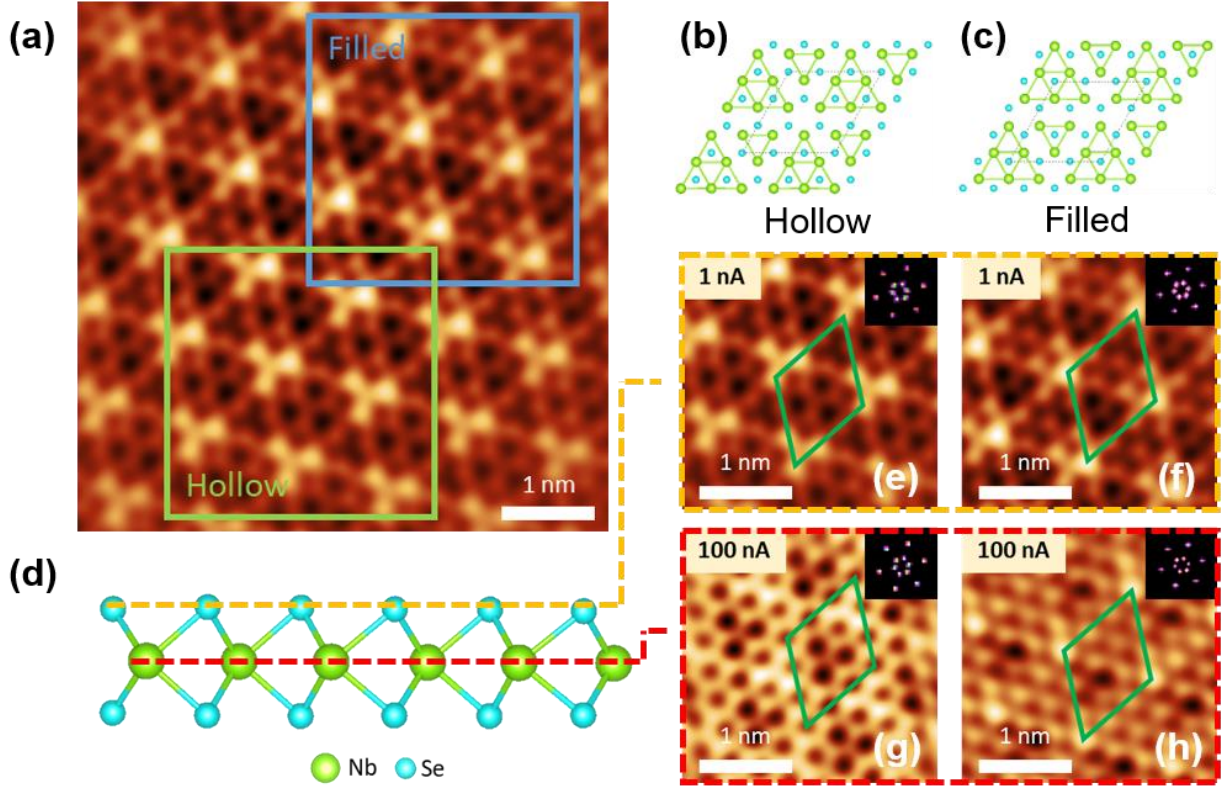

**Figure 1**. (a) The STM image of *2H*-NbSe$_2$ simultaneously shows two types of CDW phases, hollow and filled, marked with green and blue square regions, respectively. (b, c) Top view of CDW atomic models of hollow and filled phases, respectively. (f) Side view of atomic model of *2H*-NbSe$_2$. (e, f) are typical STM images of hollow and filled phases acquired at 1nA tunneling current, respectively. (g, h) are STM images acquired at the same position as (e) and (f) but at 100 nA, respectively. The insets at the right-up corner of (e-h) show FFT images of each topographies. All of the FFT images show the 1×1 Bragg peaks and the 3×3 CDW signal. All of the STM images are acquired at the sample bias of +10 mV, except (g) is obtained at -10 mV.

    **Figure 2** summarizes all STM images obtained from 0.1 nA to 100 nA at both +10 mV and -10 mV sample biases mainly for the hollow phase. Similar experiments are performed for filled phase within Supplemental Materials. It is clear that the 3×3 charge order is vivid at 0.1 nA but it is gradually getting weak at the tunneling current increases. The 3×3 charge order almost disappears at 50 nA but appears again with different appearance at 100 nA. While overlapping the atomic models with the STM images as shown in **Figure 2 (d)** to **(f)** and **Figure 2 (j)** to **(l)**, we can understand that the Se sites evolve from bright spots to dark dots from 50 nA to 100 nA. It indicates that the DOS probed by the STM tip gradually resembles the charge density at the Nb

sites at 100 nA. Thus, the DOS at Se sites decreases, showing dark dots in the STM images. In order to quantitatively understand the evolution of STM data as increasing tunneling currents, the intensity of FFT patterns corresponding to both 1×1 Bragg spots and 3×3 CDW is analyzed as shown by **Figure 2 (m)**. The change in intensity along the white arrow in **Figure 2 (m)** is shown in **Figure 2 (n)** and **(o).** It is clear that the intensity of 3×3 peaks is stronger than 1×1 at 1 nA but becomes weaker than 1×1 at 50 nA. **Figure 2 (p)** and **(q)** show the full evolutions for +10 mV and -10 mV sample biases, respectively. The relative intensity ratio at different tunneling currents are calculated by the formulae below:

$$\%_{3\times3} = I_{3\times3}/(I_{3\times3} + I_{1\times1}) \times 100\% \quad (1)$$
$$\%_{1\times1} = I_{1\times1}/(I_{3\times3} + I_{1\times1}) \times 100\% \quad (2)$$

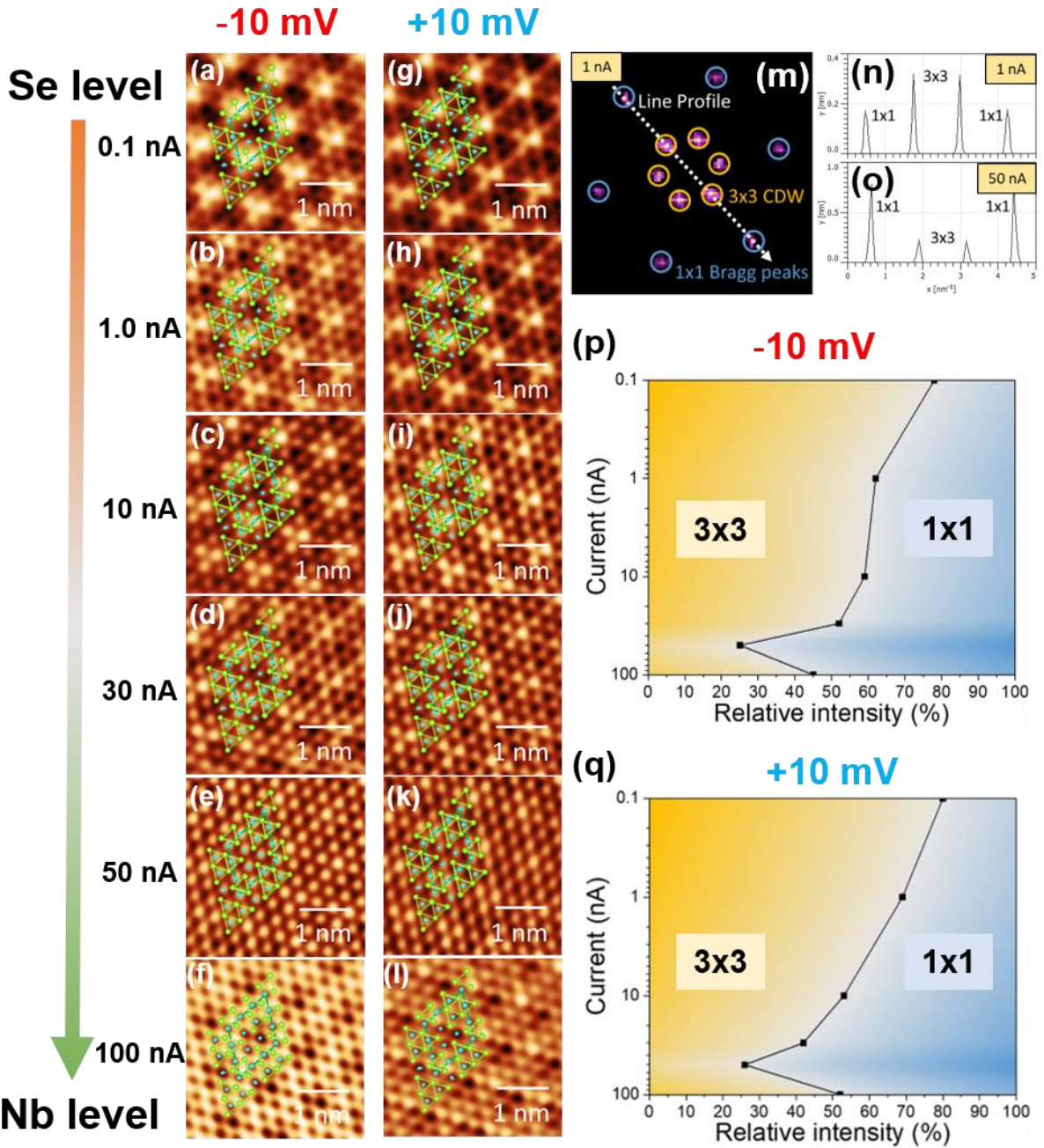

**Figure 2**. (a-f) and (g-l) are series of current-dependent STM images of hollow phase acquired from 0.1nA to 100nA at the sample bias of -10 mV and 10 mV, respectively. Atomic models of hollow phase are overlapped on (a-l) to indicate the locations of Nb and Se atoms. (m) The illustration of the method that we use to acquire the intensity of the 1×1 and the 3×3 peaks from the FFT image. (n) and (o) are profiles acquired at 1.0 nA and 50 nA, respectively, using the line-profile analytical method illustrated in (m). (p, q) The current-dependent evolution of the intensity of the 3×3 and the 1×1 signals at -10 mV and +10 mV, respectively.

To fully understand the evolution observed by current-dependent STM studies, we further compared them with the simulation by DFT calculation. **Figure 3 (a)** and **(b)** are the STM images of the hollow phase acquired at 1 nA and 100 nA at the same location. **Figure 3 (c)** and **(d)** are the respective filtered images to highlight the 3×3 charge order. By filtering the lattice information, it can be clearly seen that A site at 1 nA is the brightest but B site becomes the brightest at 100 nA while N site reveals no change. It indicates that the distribution of 2D CDW is not all the same along *c*-axis. **Figure 3 (e)** and **(f)** are the DFT-simulated STM images with the height of 5.51 Å and 3.81 Å with respect to the bottom Se atoms. Therefore, **Figure 3 (e)** is close to the outermost Se level while **Figure 3 (f)** is located near the level of middle Nb atoms. Similarly, by filtering the lattice information, **Figure 3 (g)** and **(h)** highlight the 3×3 charge order from the simulated images near Se and Nb level, respectively. The reverse of the brightness between A site and B site can be also observed in the simulated images, manifesting a special evolution of CDW in a single layer NbSe$_2$. **Figure 3 (i)** further shows all tomography scan by tuning the tunneling currents. **Figure 3 (j)** shows similar tomography scan at different height levels. The corresponding scan level is illustrated in **Figure 3 (k)**. Finally, **Figure 3 (l)** visualizes the special distribution from integrating the images by DFT simulation. The consistency between experiments and calculated results implies that a clear 3D tomography of the 3×3 CDW in between Se level and Nb level is obtained.

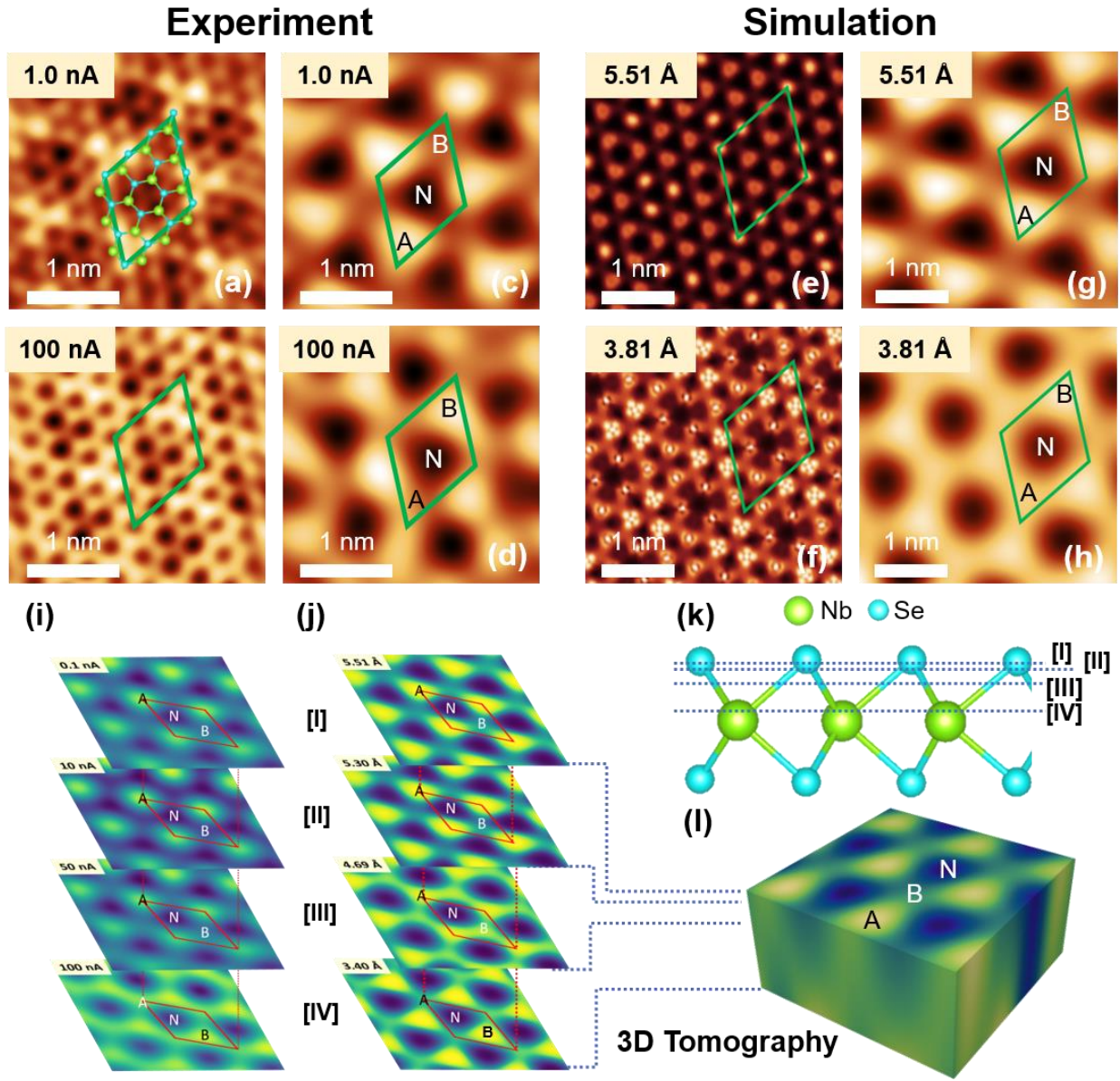

**Figure 3**. (a, b) STM images of the hollow phase at 1nA and 100 nA, respectively. (c, d) Filtered images highlight CDW signal, which are generated from (a, b), respectively. (e, f) Simulated STM images of the hollow phase at the surface (5.51 Å) and deeper position (3.81Å) of $NbSe_2$, respectively. (g, h) Fourier-filtered images of CDW-signal-only, which are acquired from (e, f), respectively. (i) Experimental tomography scan by different tunneling current. Three different sites, A, B, and N are marked for quick identification of the change in intensity. (j) Tomography scan of simulated result at different height. (k) Illustration of the scan level in the layer $NbSe_2$. (l) Real-space 3D tomography of CDW of the theoretical simulation.

To discuss the origin of this 3D spatial distribution of 3×3 CDW in NbSe$_2$, we further consider the contribution of different orbitals at Se and Nb levels. First of all, **Figure 4 (a)** to **(f)** show the images observed under different tunneling currents after filtering the 3×3 charge order. Therefore, they can be simply considered as the images showing only 1×1 lattice information. By overlapping with the atomic model, it shows that Se sites appear as bright spots at Se level and consequentially become dark spots at Nb level. Similar results can be observed in DFT calculation. **Figure 4 (g)** to **(h)** reveal the distributions of Se *p* orbitals and Nb *d* orbitals, respectively. The distribution of Se *p* orbitals matches with the filtered STM image observed at Se level where most of orbital charges is located on the Se atoms. Meanwhile, the distribution of Nb *d* orbitals shows similarities to the filtered STM image observed at Nb level. The *d* orbitals of neighboring Nb atoms overlap with each other. The calculated results further imply that the orbitals in NbSe$_2$ play a key role of rearranging the 3×3 charge order and results in the variation of its strength as observed in **Figure 2 (p)** and **(q)** and the evolution of its spatial distribution as shown in **Figure 3 (i)** and **(j)**.

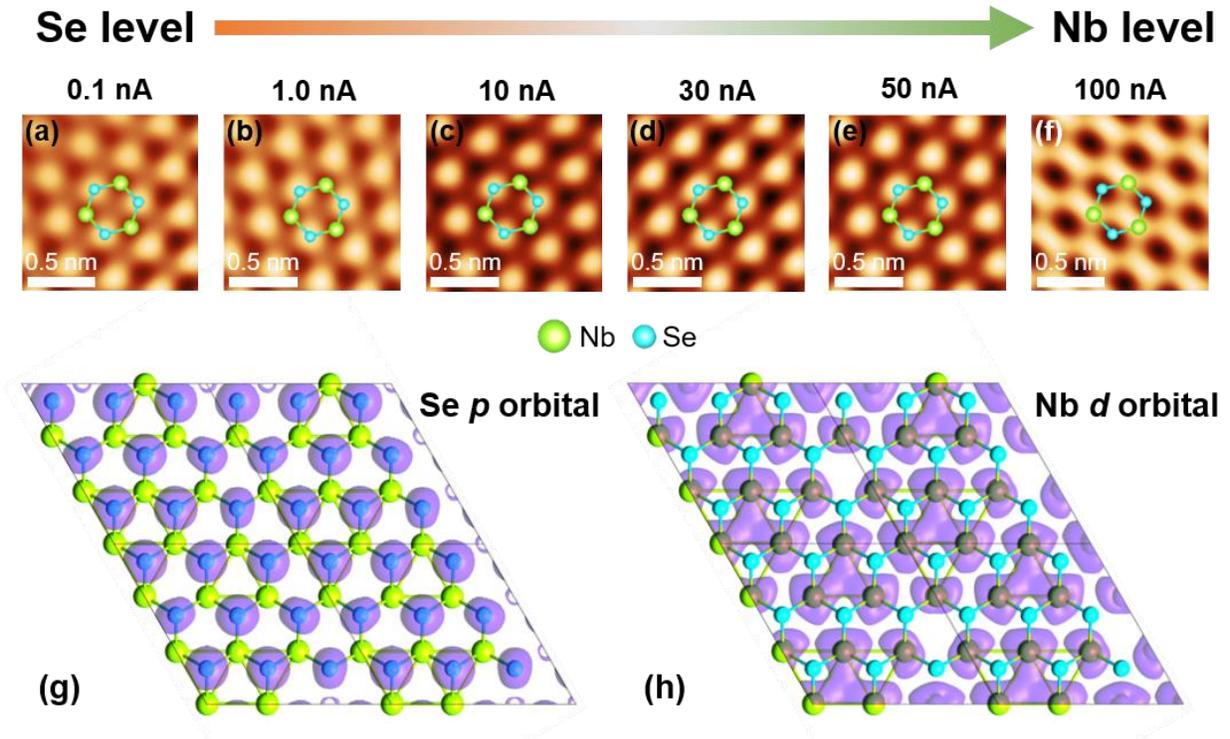

**Figure 4**. (a-f) is a series of Fourier-filtered images of 1×1-signal-only, which are acquired from 0.1nA to 100nA. Atomic models of NbSe$_2$ are overlapped on each image to indicate the location of Nb and Se atoms. (g) spatial distribution of Se *p* orbital (h) spatial distribution of Nb *d* orbital.

## Conclusion

By current-dependent STM experiments and DFT calculation, it is found that the 3×3 CDW in NbSe$_2$ exhibits a spatial distribution and a strength modulation along *c*-axis. The 3D spatial distribution is clearly visualized by a nearly tomography scan method and strongly correlated with the charges dominated by Se *p* orbitals and Nb *d* orbitals. The CDW basically modulates the orbital charges located on different height levels, resulting in different distributions. On the other hand, the strength of the 3×3 CDW is also strongly modulated by the distribution of orbital charges. Our results show that the strength of the 3×3 CDW is stronger at the level with more orbital charges. This work also gives a feedback to explain the real-space anisotropy of the superconducting gap in NbSe$_2$ and provide information to further understand similar novel quantum TMDs at diverse aspects.